\documentclass[13pt,epsf,epsfig,psfig]{article}
\usepackage{graphicx}
\usepackage{epsfig}
\usepackage{cite}
\def\J{$J/\psi$}
\def\j{J/\psi}

\def\q{q{\bar q}}
\def\Q{Q{\bar Q}}

\def\H{{\cal H}}

\def\NP{{ Nucl.\ Phys.\ }}
\def\PL{{ Phys.\ Lett.\ }}
\def\PR{{ Phys.\ Rev.\ }}

\def\ZP{{ Z.\ Phys.\ }}
\def\EPJ{{Eur.\ Phys.\ J.\ }}

\def\be{\begin{equation}}
\def\ee{\end{equation}}
\def\lsim{\raise0.3ex\hbox{$<$\kern-0.75em\raise-1.1ex\hbox{$\sim$}}}
\def\gsim{\raise0.3ex\hbox{$>$\kern-0.75em\raise-1.1ex\hbox{$\sim$}}}

\begin{document}

\parindent=0pt 

%4.\ 2.\ 2009

%\vskip1cm

\centerline{{\Large \bf Heavy Quark Interactions and Quarkonium 
Binding}\footnote{Based on joint work with O.\ Kaczmarek and F.\ Karsch}}

\vskip0.5cm

\centerline{\bf Helmut Satz}

\bigskip

\centerline{Fakult\"at f\"ur Physik, Universit\"at Bielefeld}

\centerline{D-33501 Bielefeld, Germany}

\vskip1cm

\centerline{\bf Abstract}

\medskip

We consider heavy quark interactions in quenched and unquenched 
lattice QCD. In a region just above the 
deconfinement point, non-Abelian gluon polarization leads to a strong 
increase in the binding. Comparing quark-antiquark and quark-quark 
interaction, the dependence of the binding on the separation distance 
$r$ is found to be the same for the colorless singlet $\Q$ and the colored 
anti-triplet $QQ$ state. In a potential model description of in-medium 
\J~behavior, this enhancement of the binding leads to a survival up to 
temperatures of 1.5 $T_c$ or higher; it could also result in \J~flow.

\section{Introduction}

\begin{figure}[htb]
{\epsfig{file=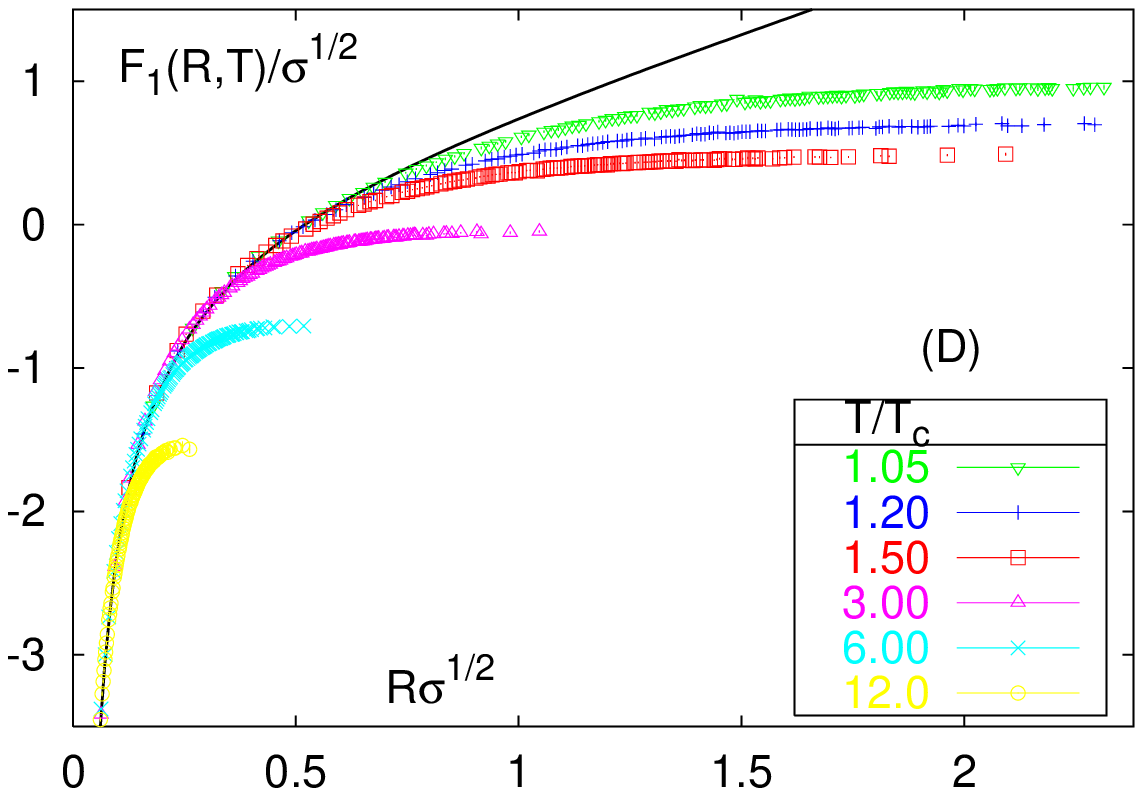,width=7cm,height=5.5cm}}
\hfill{\epsfig{file=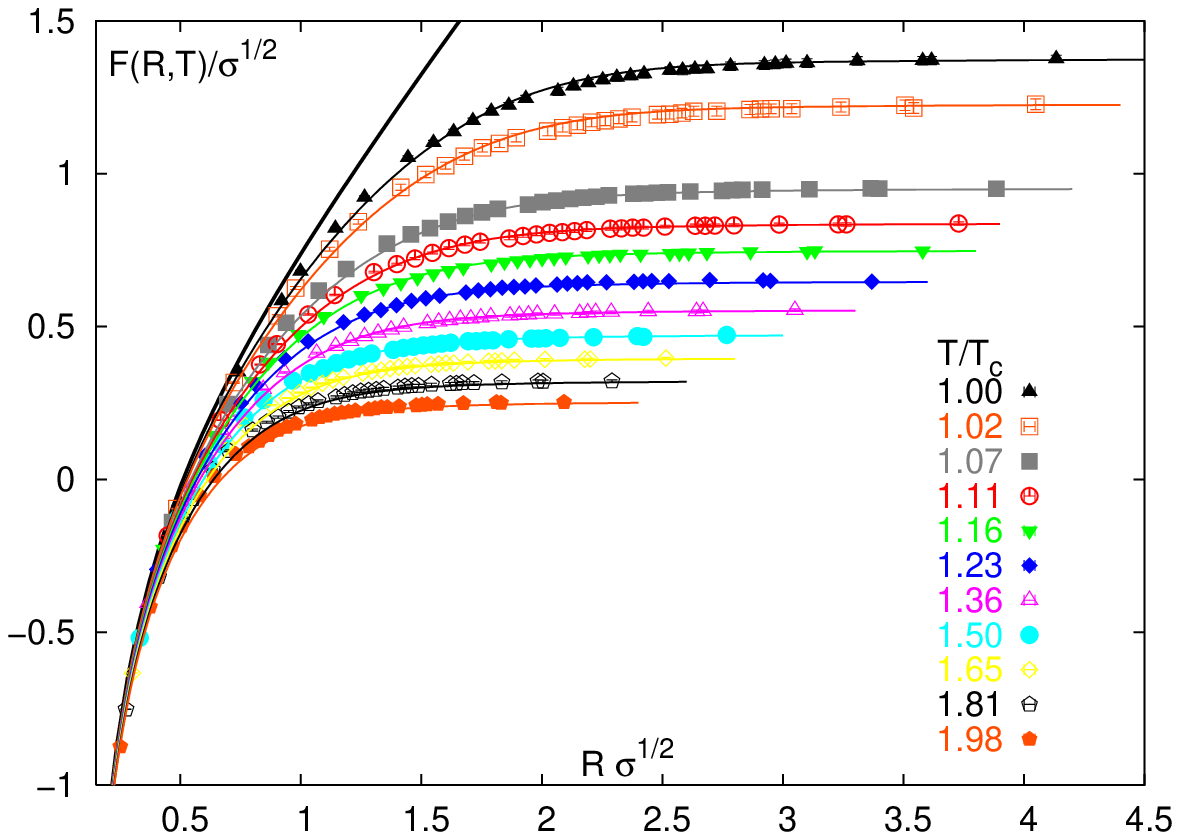,width=8cm,height=5.5cm}\hskip1cm}
\vskip-0.3cm
\hskip3.7cm (a) \hskip7.5cm (b)
\vskip0.2cm
\caption{Free energy difference for a color-singlet 
$\Q$ pair as function of $r$,
(a) for quenched and (b) for two-flavor QCD \cite{KFPZ}.}
\label{F-r}
\end{figure}

The interaction of a heavy quark-antiquark ($\Q$) pair in strongly 
interacting matter has been studied in finite temperature lattice QCD 
in the quenched approximation as well as for the cases of two light and 
two light plus one heavy quark species \cite{KFPZ,P-P,KZ,KZ-lat05}.
In all these 
studies, one obtains the difference $F(r,T)$ between the color singlet 
free energy with and without the heavy quark pair, as function of the 
temperature $T$ of the medium and the separation distance $r$ of $Q$ 
and $\bar Q$. Schematically, we can write
\be
F_0(T) =-T~\! \ln \int d\Gamma \exp\{-\H_0/T\}
\label{Fohne}
\ee
and
\be
F_Q(r,T) =-T~\! \ln \int d\Gamma \exp\{-\H_Q(r)/T\}
\label{Fmit}
\ee
for the free energies in question; here the Hamiltonian $\H_0$ describes 
the interacting quark-gluon plasma alone, and $\H_Q(r)$ that of the plasma 
containing the static $\Q$ pair; the integral $\int d\Gamma$ denotes the
grand canonical phase space integration and summation. The color singlet
free energy 
difference provided by lattice studies is then defined as
\be
F(r,T) = F_Q(r,T) - F_0(T).
\label{Fdiff}
\ee
In Fig.\ \ref{F-r}, we show the $r$-dependence of $F(r,T)$ at different 
temperatures above the deconfinement point, i.e., for $T> T_c$. The vacuum 
form (for $T=0$) also shown here is the Cornell potential \cite{Cornell}
\be
 V(r) = \sigma ~r - {\alpha \over r},
\label{cornell}
\ee
defined in terms of the string tension $\sigma$ and the Coulomb coupling 
$\alpha \simeq \pi/12$. For very small $r \ll T^{-1}$, the small color
singlet pair is neither seen nor affected by the medium, so that the 
interaction is specified by the $T$-independent running coupling 
$\alpha(r)$, which for larger $r$ saturates at the canonical string value
$\pi/12$. On the other hand, at very high temperatures and comparatively 
large $r \gg T^{-1}$, in the perturbative limit, we expect an 
$r$-independent running coupling $\alpha(T)$. 

\medskip

In full QCD below $T_c$, the presence of light quarks ($q$) leads to 
$\q$ pair production and string breaking; from quarkonium studies, the 
string breaking energy at $T=0$ is estimated to be about 1.0 - 1.2 GeV.

\medskip

\medskip

In the quenched case below $T_c$, the free energy diverges in the large
distance limit, with a temperature-dependent string tension $\sigma(T)$
which vanishes at $T=T_c$. Temperatures above $T_c$, however, as seen in 
Fig.\ \ref{F-r}a, also lead to a finite large-distance limit, with a 
temperature-dependence which is very similar to that found in full QCD. 
Since here there are no light $\q$ pairs to provide string-breaking,
the large distance behavior in the quenched case must arise from gluon 
screening effects, i.e., it is of a non-Abelian origin. We therefore
expect that also in full QCD gluonic screening plays a crucial role.

\medskip

\begin{figure}[htb]
{\epsfig{file=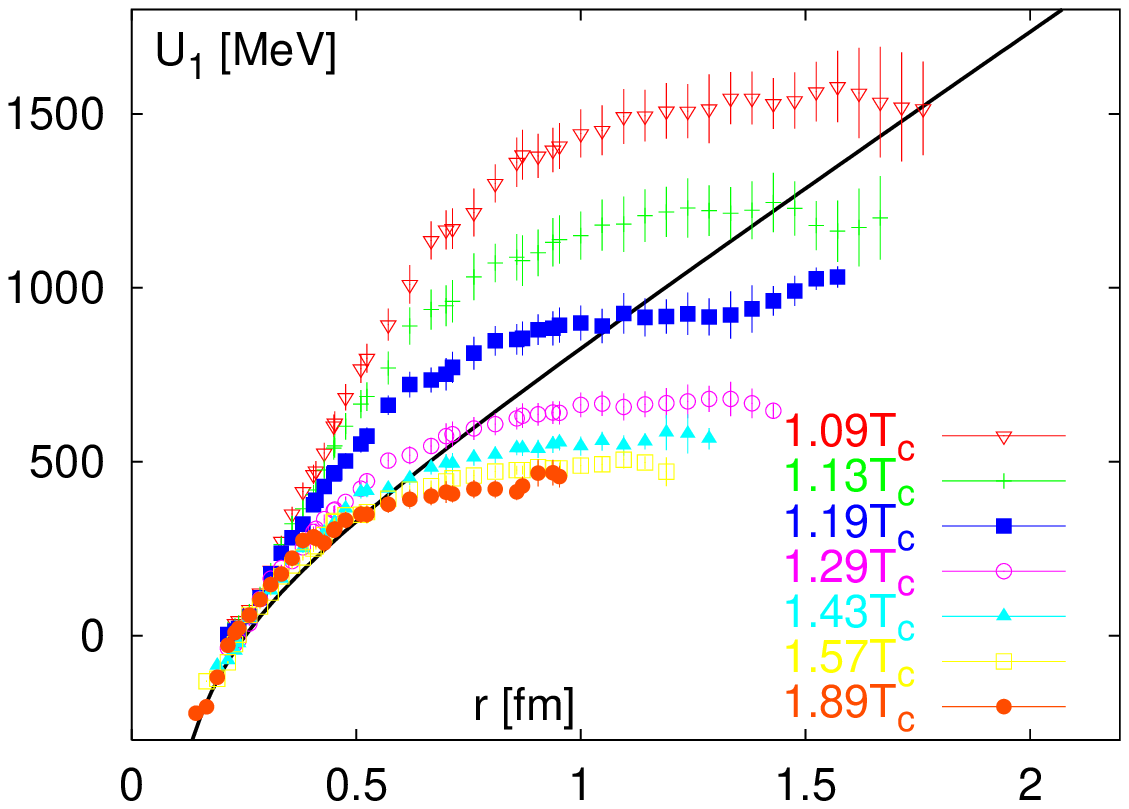,width=8cm,height=6cm}}
\hfill{\epsfig{file=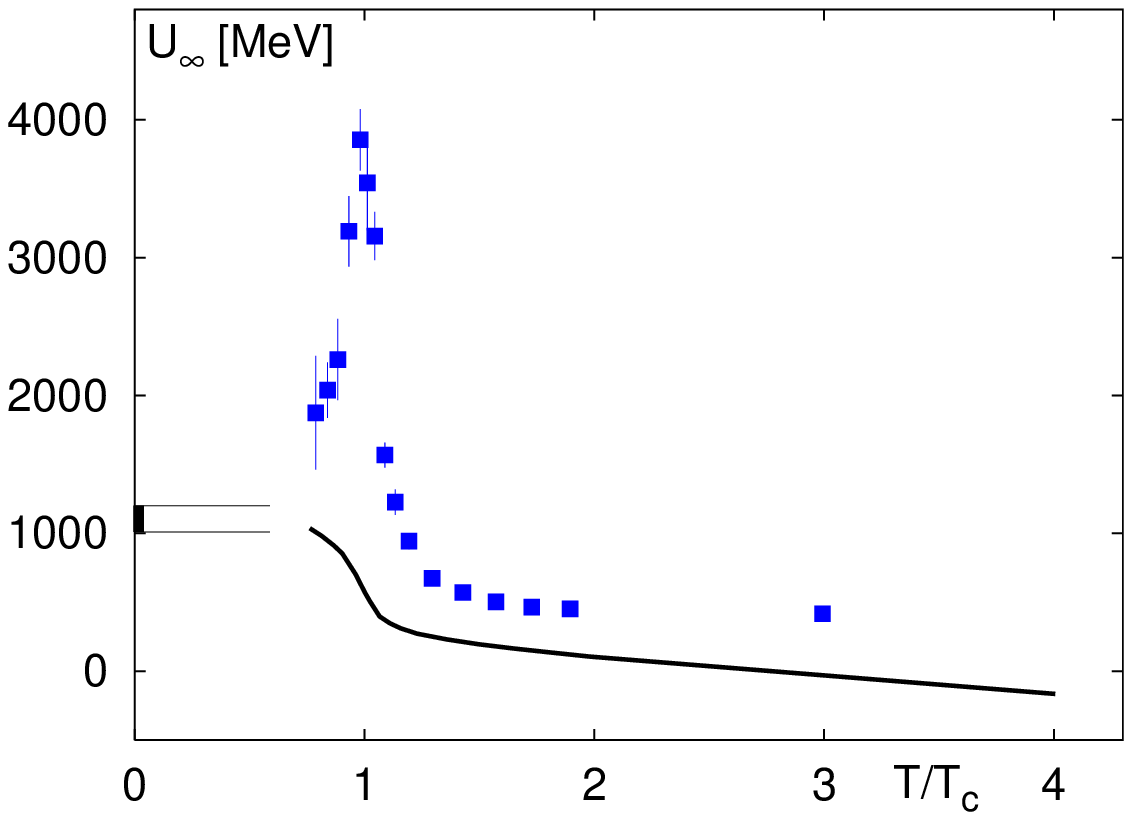,width=8cm,height=6cm}\hskip1cm}
\vskip-0.1cm
\hskip3.7cm (a) \hskip7.5cm (b)
\vskip0.2cm
\caption{Internal energy difference for a color-singlet $\Q$ pair in 
two-flavor QCD, (a) as function of $r$ and (b) for $r\to \infty$, 
compared to the corresponding free energy (solid line) \cite{KZ-lat05}.}
\label{U}
\end{figure}

From $F(r,T)$ one obtains through standard thermodynamic relations 
the corresponding difference for the internal energy $U(r,T)$, 
\be
U(r,T) = -T^2 \left({\partial [F(r,T)/T] \over \partial T}\right)
= F(r,T) - T\left( {\partial F(r,T) \over \partial T} \right)
= F(r,T) + T~S(r,T), 
\label{u}
\ee
and for the entropy $S(r,T)$,
\be
~~S(r,T) = -\left({\partial F(r,T) \over \partial T}\right). 
\label{s}
\ee
The internal energy behavior is shown in Fig. \ref{U} \cite{KZ,KZ-lat05}. 
From eqs. (\ref{Fohne}), (\ref{Fdiff}) and (\ref{u}), we obtain
\be
U(r,T) = \langle \H_Q(r,T) \rangle - \langle \H_0(T) \rangle,
\label{Hdiff}
\ee
which for a static $\Q$ pair, with no kinetic energy, measures the change 
in potential energy due to the introduction of the pair. It is seen in
Fig.\ \ref{U}a that just above the deconfinement
point, the potential is much stronger than at $T=0$. In the following,
we want to study the origin of this increase. 

\medskip

\medskip

Consider first the underlying medium, a plasma of unbound (but interacting) 
gluons and (for full QCD) quarks at temperature $T$. We had seen that
gluon screening plays a decisive role in the medium, and when we speak 
of constituents, this should be kept in mind. Any interaction is mediated 
by gluons, and when the gluonic interaction range is reduced through color 
screening, this holds for quark interactions as well. We denote by $n(T)$ 
the average density of constituents, so that $d=n^{-1/3}$ is the average 
separation distance between adjacent color charges. The interaction range 
for the constituents is given by the correlation function,
\be
f(r) \sim \exp\{-r/\xi(T)\},
\label{cor}
\ee
which 
measures the interaction strength between constituents separated by a 
spatial distance $r$. The correlation length $\xi(T)$ is a basic property 
of the medium, and it will diverge for $T\to T_c$ in case of a second order 
confinement/deconfinement transition. In the high temperature limit, we 
expect a non-interacting gas and hence $\xi \to 0$.

\medskip

We now add a static $\Q$ pair to our system. As long as the $Q$ and the 
$\bar Q$ are sufficiently far apart, their color charges are effectively 
screened and asymptotically, they do not interact with each other.
Nevertheless, the $Q$ as well as the $\bar Q$ individually interact
with the medium, and around each charge, this will lead to polarization 
effects in the region where its presence is felt. Hence in a sphere of 
average radius $\xi(T)$ around the $Q$ (and similarly around the $\bar Q$),
there will be an excess of color compared to the state of the medium before 
insertion of the static charges. The polarization screens each static 
charge, so that its strength is reduced the further away we measure it. 
This evidently is a non-Abelian effect, since an electric charge cannot 
be screened by photons, and in quenched QED the $\Q$ interaction would 
remain purely Coulombic even in the large distance limit. 
When the $\Q$ separation distance $r$ is reduced, eventually interaction 
between the charges becomes significant. This interaction initially has a 
Debye-screened Coulomb form,
with $\alpha_{\rm eff} \sim \alpha~\!\exp\{-\mu r\}$ as the effective
coupling, where $\mu$ denotes the screening mass. It has to be emphasized 
that such a picture makes sense only for $r \gg \xi$, so that we really 
have two well separated polarization spheres, each of which contains a 
static charge surrounded by many medium charges, with an overall excess 
of the opposite charge.
  
\medskip

When the heavy quarks are brought still closer to each other, the 
polarization spheres begin to overlap \cite{KZ-lat05}.
The overlapping spheres continue to attract each other, but this
interaction now has two components and can no longer be described
by a pure Debye-screened potential. On one hand, there is the direct 
interaction between the $Q$ and the $\bar Q$, which in the limit of 
small $r$ becomes just the Coulombic vacuum form $\alpha/r$. On the 
other hand, the charged constituents inside the overlapping polarization 
spheres also attract each other directly, with a strength determined by 
the deviation of the region from the state without a $\Q$ pair.
Once the spheres no longer overlap, this interaction becomes by 
Gauss' law just the Debye-screened Coulomb form with a reduced effective
charge.   

\medskip

In order to determine the behavior of $\Q$ binding in a deconfined medium,
we shall first consider the large and small distance limits of the 
thermodynamics potentials.

\section{The Large Distance Limit}

In the limit of large $\Q$ separation, for $r \to \infty$, we have 
two fully screened and hence non-interacting charges. Nevertheless, the 
thermodynamic potential differences $F,U$ and $S$ do not vanish: 
they specify the effect of the interaction of each of the two independent 
charges with the medium. This is not related to the interaction of $Q$ 
and $\bar Q$, 
as is easily seen by considering the corresponding quantities for a pair 
of heavy quarks (a ``diquark''). Lattice calculations \cite{Doering,GHK}
have in fact shown 
that for $r \to \infty$ one obtains
\be
F^{(1)}_{\Q}(T) = F^{(\bar 3)}_{QQ}(T) \equiv 2 F_3(T),
\label{larged}
\ee
where $F^{(1)}$ corrresponds to the color singlet state of the $\Q$
and $F^{(\bar 3)}$ to the anti-triplet state of the diquark. Corresponding 
relations hold for $U(T)$ and $S(T)$. Hence $F_3(T)$ as defined
above indeed specifies the free energy associated to a single static
(triplet or anti-triplet) color charge in a medium of temperature $T$; 
$U_3(T)$ and $S_3(T)$ denote the corresponding internal energy and 
entropy. The relation
\be
U_3(T) = F_3(T) + T~\!S_3(T)
\label{core-medium}
\ee
can then be interpreted as the sum of the energy shift $F_3(T)$ due to 
the interaction of the static $Q$ with the constituents of the medium, 
and the energy shift $T~\!S_3(T)$ due to the interaction between the 
constituents of the cloud, 
with $S_3(T)$ specifying their change in number and $T$ their energy. It 
is clear that $U_3(T)$ is not due to any interaction between the static 
$Q$ and its partner $\bar Q$, and it will thus not affect the binding
of the heavy quark pair. 

\section{The Short Distance Limit}

For the $\Q$ system, we now have to distinguish between the attractive 
color singlet and the repulsive color octet state. In the case of the 
singlet, when $r$ become sufficiently small compared to the average 
separation of charges in the medium, the $\Q$ pair constitutes for 
the medium a color-neutral entitity and is not ``seen'' by its constituents. 
Hence for $r \ll T^{-1}$, the differences $F^{(1)}_{\Q}(r,T)$ and 
$U^{(1)}_{\Q}(r,T)$ measured in lattice studies are simply due to the 
perturbative interaction of the two static charges, with no medium 
effects. In the short distance regime we thus have the 
temperature-independent form
\be
F^{(1)}_{\Q}(r,T) = U^{(1)}_{\Q}(r,T) = - {4\over 3} {\alpha(r) \over r},
\label{short-sing}
\ee
where $\alpha(r)$ is the $r$-dependent running coupling and $4/3$ is the
$SU(3)$ color Casimir coefficient for $3 \times \bar 3 \to 1$. 

\medskip

\begin{figure}[htb]
\centerline{\epsfig{file=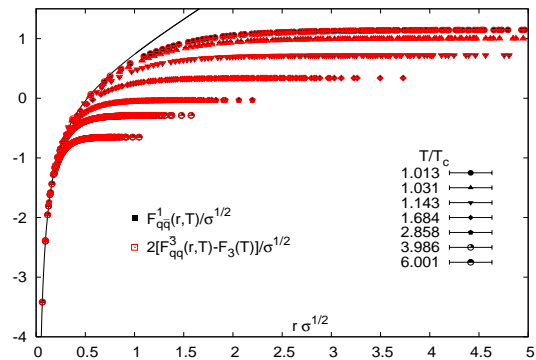,width=7.5cm,height=5cm}}
\vskip0.2cm
\caption{Free energy of a singlet $\Q$ and of an
anti-triplet $QQ$ \cite{Doering}, see eq.\ (\ref{sing-anti})
\cite{Doering}.}
\label{sing-anti-inter}
\end{figure}

For the attractive anti-triplet case of the diquark state, the perturbative 
form of the direct $QQ$ interaction gives $1/2$ that of the singlet $\Q$,
as determined by color $SU(3)$ Casimir coefficient $3 \times 3 \to
\bar 3$. Now, however, 
the small overall system is still colored and hence is seen by the medium 
as a single point-like color charge. It will thus again lead to a 
polarization cloud, which for a triplet or anti-triplet just is $F_3(T)$. 
As a result, we have in the short distance regime the free 
energy \cite{Doering}
\be
F^{(\bar 3)}_{QQ}(r,T) \simeq - {2 \over 3} {\alpha(r) \over r} +  F_3(T).
\label{short-anti}
\ee
Comparing eqs.\ (\ref{short-sing}) and (\ref{short-anti}), we expect that 
\be
F^{(1)}_{\Q}(r,T) \simeq 2[ F^{(\bar 3)}_{QQ}(r,T) - F_3(T)]
\label{sing-anti}
\ee 
should be satisfied in the small distance limit. Using eq.\ (\ref{larged})
we see that it also holds at large distance, and as seen in Fig.\ 
\ref{sing-anti-inter}, it does so for $T>T_c$ even at intermediate 
separation distances. 

\medskip

\medskip

We thus find a general pattern of behavior for the attractive 
case (colorless singlet $\Q$ and colored anti-triplet $QQ$).
The only caveat is that while the singlet $\Q$ becomes simply
Coulomb-like at short distance, for the attractive $QQ$ antitriplet
state a polarization cloud remains even in the short distance limit.
Once this is taken into account, the remaining interaction appears to be
quite insensitive to whether we consider a $\Q$ or a $QQ$ system.

\section{The Intermediate Separation Regime}

\begin{figure}[htb]
{\epsfig{file=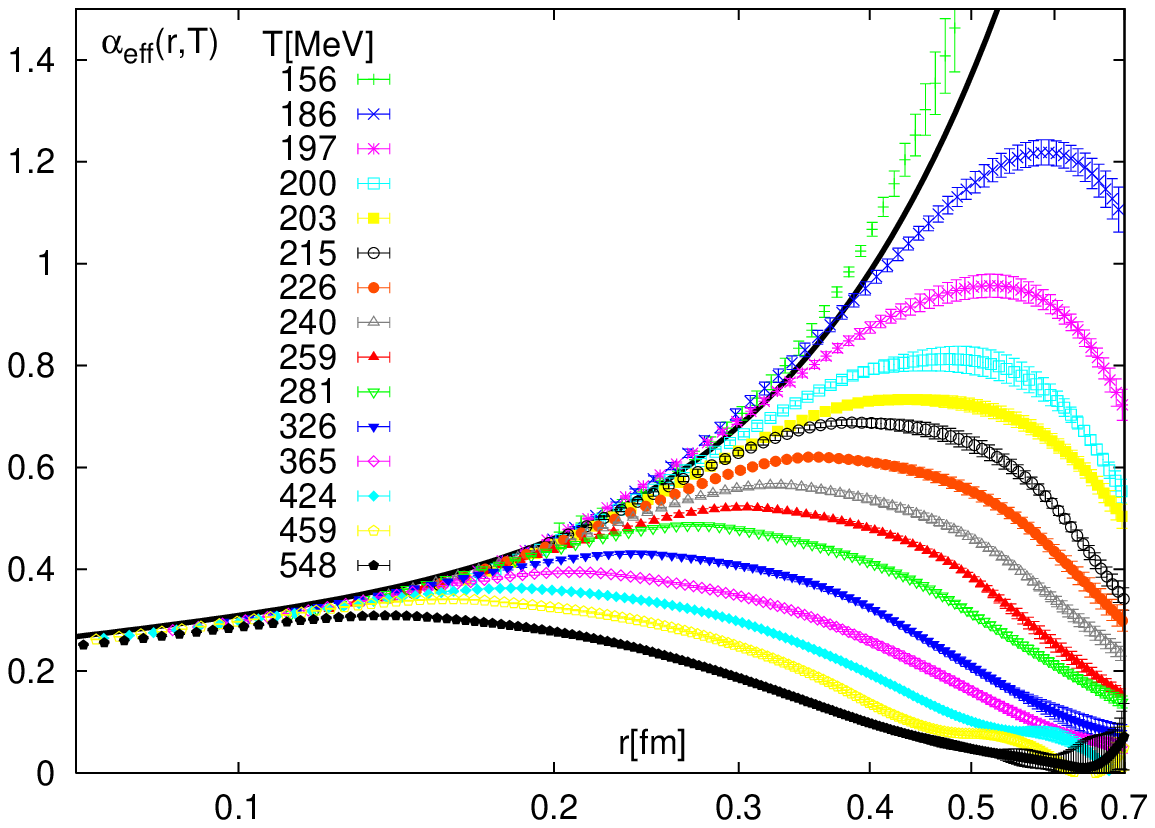,width=7.5cm,height=5.5cm}}
\hfill{\epsfig{file=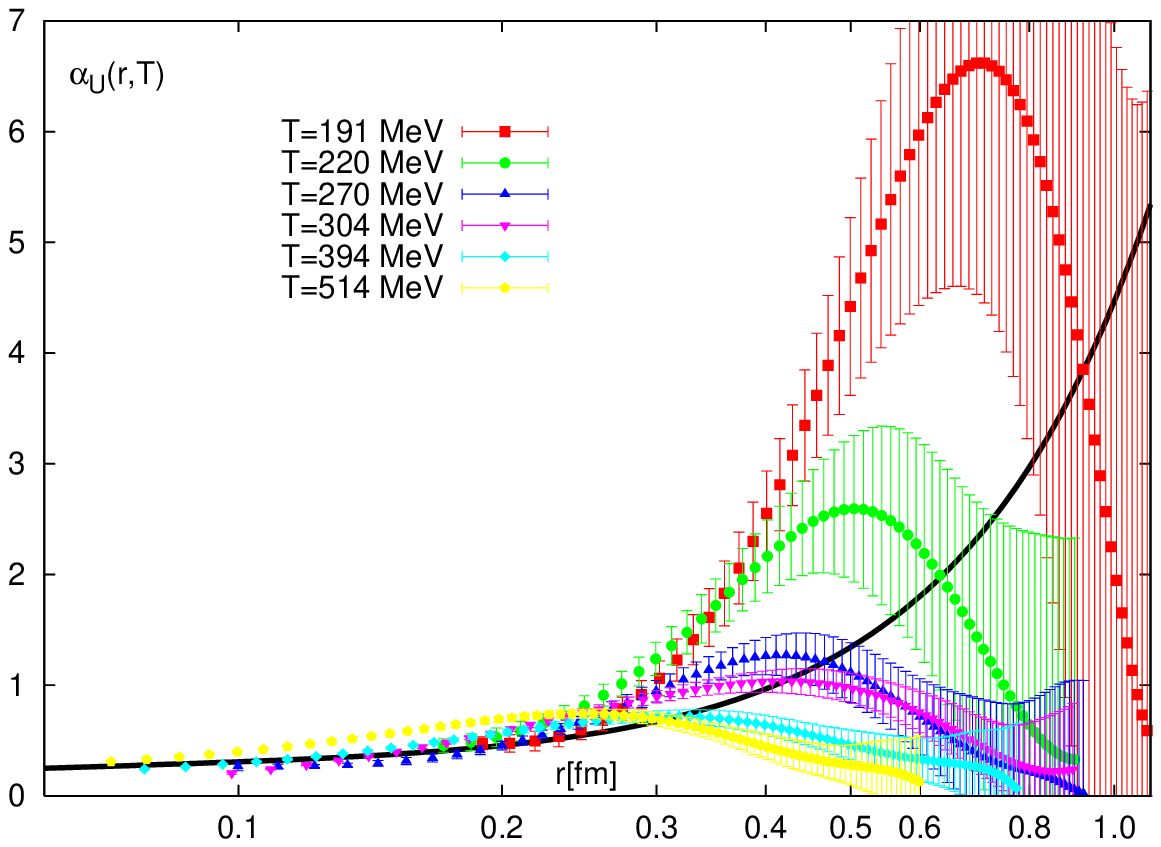,width=7.5cm,height=5.5cm}\hskip1cm}
\vskip-0.1cm
\hskip3.7cm (a) \hskip7.5cm (b)
\vskip0.2cm
\caption{Effective couplings from (a) free and (b) internal energy
as function of $r$ \cite{Okacz}.}
\label{al-FU}
\end{figure}

To study the behavior at finite $r$, we consider the effective couplings 
$\alpha_F$ and $\alpha_U$ for the singlet $\Q$ state, with
\be
\alpha_F(r,T) = {3\over 4} r^2 \left({\partial F(r,T) \over 
\partial r}\right),
\label{alpha-F}
\ee
and a corresponding relation for $U(r,T)$ in place of $F(r,T)$. For 
sufficiently small $r$, there are no medium effects and hence $\alpha_U = 
\alpha_F$. Since $S(r,T)$ becomes constant for large $r$, they also become 
equal in that limit. The behavior of $\alpha_F(r,T)$ for the singlet $\Q$ 
system, as found in lattice QCD, is illustrated in Fig.\ \ref{al-FU}a for 
a range of temperatures $T > T_c$ \cite{Okacz}. It is seen that 
$\alpha_F(r,T)$ follows the vacuum form (\ref{cornell}) up to the 
point where Debye screening begins to set in and eventually makes it vanish.
In the range of $r$ values of interest here,
this behavior disagrees strongly with a Yukawa-like Debye-screened form 
$F_1(r,T) \sim \alpha(T) r^{-1} \exp\{-r/r_D(T)\}$, with $r_D(T)$ as 
color-screening radius; the actual form has been studied in detail 
in \cite{Digal}. As a result, models based on a Yukawa form \cite{Brau}
cannot correctly
reproduce the free energy data in the crucial range of $r$ and $T$; 
either $\alpha$, or $m_D$, or both must depend on $r$ as well as on $T$.   
In Fig.\ \ref{al-FU}b, we also show the corresponding behavior of
$\alpha_U(r,T)$. As argued above, the two couplings agree at fixed $T$ 
in the small as well as in the large distance limits. In the intermediate
$r$-range, the internal energy coupling considerably exceeds the
vacuum Cornell form as well as that obtained from the free energy. The 
reason for this excess is that the internal energy difference contains 
not only the $\Q$ interaction, but also the effect of the interaction 
of the charges in the polarization clouds. 
With increasing temperature, the correlation length and hence the
size of the polarization clouds decreases, so that the difference 
between the two couplings also decreases. 

\medskip

\medskip

The difference between the values of $U(r,T)$ and $F(r,T)$ obtained in
lattice studies thus becomes clear. The internal energy measures the
binding potential between the two heavy quarks, that between each quark 
and the constituents of its polarization cloud, and that between the
constituents of the overlapping polarization clouds. The
free energy, on the other hand, measures only the potential between the
``bare'' heavy quarks, modified by the screening effects due to
the polarization clouds. In the large distance
limit, the heavy quark potential vanishes, so that $F$ approaches with
increasing temperature the entropy change $-T~\!S(T)$; since the
entropy change becomes very weakly (logarithmic) temperature-dependent 
for large $T$, this change grows almost linearly with $T$. For 
$r \to \infty$, $U(r,T)$ approaches
the twice interaction energy between a heavy quark and its cloud
constituents; if the correlation length and hence the cloud size 
vanish in this limit, $U(\infty,T)$ is also expected to vanish.
In the short distance limit, the $\Q$ screens itself, so that both
$F$ and $U$ measure only the direct $\Q$ interaction.   

\medskip

Stated in other words, the attractive interaction between two heavy quarks 
inside a deconfined medium has two distinct components: the direct 
Coulombic $\Q$ interaction, and the non-Abelian
interaction between the constituents of the
polarization clouds which the static quarks acquire in the medium. At 
large distances, this ``dressing'' simply reduces the effective charge 
through screening. But at intermediate distances, when the dressings 
overlap, they provide a strong additional attractive interaction. 
To corroborate this quantitatively, one would need to show that
the color-averaged gluon-gluon interaction is attractive. For static
quark sources of different quantum number, including adjoint octet
quarks, such an attraction has been shown \cite{GHK}. For
gluons, it is not evident how similar studies could be carried out.
The general effect, however, is 
conceptually similar to the energy loss of a fast parton passing 
through a deconfined medium: it loses energy through direct 
(bremsstrahlung) interactions with the medium \cite{G-W}, but much more 
through the indirect interactions of its gluon cloud with the medium 
\cite{BDPS}. 

\medskip

It should be emphasized that the non-perturbative interaction observed
in the intermediate $r$ regime occurs (apart from the differences in the
short distance limit) for a colored anti-triplet $QQ$ state in just the same
functional form as it does for the colorless $\Q$ singlet.  

\section{Potential Model Studies}

These considerations can throw some light on the problem of extracting 
from lattice results the temperature-dependent $\Q$ binding potential 
$V(r,T)$, to be used in a two-body Schr\"odinger equation, 
\be
\left\{2m_c -{1\over m_c}\nabla^2 + V(r,T)\right\} \Phi_i(r) = M_i \Phi_i(r),
\label{schroedinger}
\ee
in order to study charmonium dissociation. Here $m_c$ denotes the charm
quark mass, $M_i$ that of charmonium state $i$. First studies had used the 
color-averaged free energy for $V(r,T)$ \cite{digal}, but subsequently 
lattice results for the color-singlet free energy became available, 
leading to somewhat stronger binding. Eventually it was argued that 
the correct potential is given by the internal energy $U(r,T)$. The 
proposals of the last years now cover the whole range, 
with the general form $xU(r,T) + (1-x)F(r,T)$, where $0 < x <1$
\cite{digal,SZ,Wong,Alberico,HSjpg}. As is 
evident from what was said here, the effective binding increases with $x$. 
The internal energy
$U(r,T)$, as the expectation value of the difference of the Hamiltonians 
with and without the pair, includes the indirect ``cloud-cloud'' binding 
in addition to the direct $\Q$ interaction and hence provides a stronger
binding. We believe that the indirect binding cannot be neglected, so that 
the correct form of the potential should indeed be $U(r,T)$. 

\medskip

\begin{figure}[htb]
\centerline{\epsfig{file=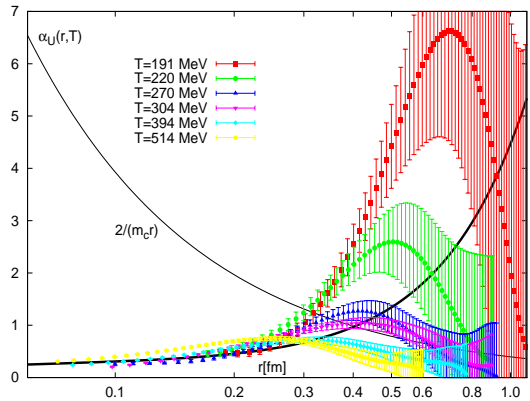,width=7.5cm,height=5.5cm}}
\caption{Semi-classical picture of \J~binding and dissociation
ranges}
\label{cross}
\end{figure}

To clarify the situation further, let us consider the semi-classical
limit of eq.\ (\ref{schroedinger}) applied to the case of \J~dissociation,
\be 
2m_c +{p^2 \over m_c
} + U(r,T) =
2m_c +{c \over m_c r^2} + U(r,T) = M_{\j}(r,T),
\label{semi}
\ee
where the minimum of the energy
\be
E(r,T) = {c \over m_c r^2} + U(r,T)
\label{energy}
\ee
as function of $r$ determines the ground state mass $M_{\j}$ of the \J.
The constant $c$ arises from the uncertainty relation $p^2 r^2 \simeq c$ and 
depends on the form of the binding potential. We determine it by requiring
the correct \J~mass at $T=0$. From
\be
{\partial E(r,T) \over \partial r} = 0 
\label{min}
\ee
we obtain
\be
c= {mr^3 \over 2} \left[ \sigma + {\alpha \over r^2}\right]
\label{constant}
\ee
which leads to 
\be
M_{\j}(r,T\!=\!0)= 
2m_c + {3\over 2}\sigma r - {\alpha \over 2r} = 3.1~{\rm GeV}.
\label{psimass}
\ee 
With $m_c=1.3$ GeV, $\sigma=0.2$ GeV$^2$, and $\alpha=\pi/12$, this
results in $c=1.56$ and for the radius of the \J~the value 
$r(T=0)\simeq 0.42$ fm, which agrees very well with that obtained in the
corresponding quantum-mechanical study \cite{KMS}. The minimization
requirement (\ref{min}) for finite $T$ leads to
\be
{2 c\over m_c r} = r^2 {\partial U \over \partial r} = {4\over 3} 
\alpha_U(r,T),
\label{Tmin}
\ee
where $\alpha_U(r,T)$ is the effective coupling determined above
(see Fig.\ \ref{al-FU}). In Fig.\ {\ref{cross}}, we solve this equation 
graphically. It is seen that up to some temperature value 
$T_{dis} \simeq 1.5 ~T_c$, $\alpha_U(r,T)$ attains a peak large 
enough to intersect the kinetic term $c /m_c r$, so that there is 
a minimum in the energy and hence a bound state. For $T > T_{dis}$, 
this is no longer the case, the energy decreases monotonically with 
$r$ and there is no more bound state. The radius of the surviving \J~is 
seen to vary very little with temperature; it remains in the range 
$0.3 - 0.45$ fm up to $T_{diss}$, where the bound state disappears.

\medskip

\medskip

The potential $U(r,T)$ used in the Schr\"odinger equation has the
correct $T$-independent form for small $r$, and enhanced attraction
in the intermediate $r$ range up to $T_{diss}$. At large $r$, it leads
to the $r$-independent value $2 U_3(T)$, which, as we saw above,
has little effect on the strength of the binding. Its value does, 
however, specify the binding energy as the gap between bound and 
free heavy quarks. With
\be
\Delta~\!E(T) = 2[m_c + U_3(T)] - M_{\j}(T)
= 2 U_3(T) - {c \over mr_0^2} - U(r_0,T)
\label{BE}
\ee
as binding energy, with $r_0$ determined by the minimization condition
(\ref{Tmin}), we obtain at the dissociation point a value of
$\Delta E(T_{diss}) \simeq 0.2$ GeV. 

\medskip

For $T<T_{dis}$, the kinetic curve in Fig.\ \ref{cross} crosses the potential
curve at two $r$ values: the first crossing specifies the minimum of the 
energy and hence the \J~bound state radius, while the difference between
the second and the first provides a measure of the thickness $\Delta r$ 
of the potential wall. As $\Delta r \to 0$, quantum effects (tunneling) 
will become more and more likely and thus provide a quantum-mechanical 
possibility of the dissociation even below $T_{dis}$, as well as a finite 
binding probability even above $T_{dis}$. 

\section{Charmonium Flow}

Our arguments suggest that charmonium binding in a hot quark-gluon plasma
is to a large extent due to a binding between the clouds of the medium 
surrounding the heavy quarks. If this medium is experiencing an overall
flow, the motion of the clouds will be transmitted to the $\Q$ core and
cause a drag leading to charmonium flow. Similarly, the isolated
heavy quarks (for $r \to \infty$) will experience the drag of their
polarization clouds, and at hadronization lead to flow of open charm
mesons. An observation of \J~flow in heavy ion experiments should
therefore not be interpreted as evidence for primary \J~dissociation,
followed by regeneration due to binding of charm constituents from
different collisions. The cloud-cloud binding discussed here can
transfer any medium motion also to a primary $\Q$ pair.

\section{Outlook}

Understanding the relation between lattice and potential studies of 
quarkonium binding is obviously an essential step in solving the
in-medium behavior of quarkonia. What we have presented here is
certainly not a final answer. It is only meant to recall 
\begin{itemize}
\vspace*{-0.2cm}
\item{that the heavy quark interaction in a quark-gluon plasma is not
a simple two-component problem, but that on the other hand,}
\vspace*{-0.2cm}
\item{this does not rule out a description in terms of a suitably
formulated potential theory.}  
\vspace*{-0.2cm}    
\end{itemize}
Recent analytical work \cite{Nora,Laine,OP,Blai} may provide a key to 
further developments in this direction.

\end{document}